\documentclass[twocolumn,showpacs,aps,pra]{revtex4-1}

\usepackage{amsmath,amsfonts,amssymb}
\usepackage{wrapfig}
\usepackage{graphicx}
\usepackage{bbm}
\usepackage{tabularx}
\usepackage{color}

\begin{document}

\title{Certifying single-system steering for quantum information processing}\date{\today}

\author{Che-Ming Li$^{1,2}$}
\email{cmli@mail.ncku.edu.tw}
\author{Yueh-Nan Chen$^{3,4}$}
\author{Neill Lambert$^{2}$}
\author{Ching-Yi Chiu$^{1}$}
\author{Franco Nori$^{2,5}$}

\affiliation{$^1$Department of Engineering Science, National Cheng Kung University, Tainan 701, Taiwan}
\affiliation{$^2$CEMS, RIKEN, Wako-shi, Saitama 351-0198, Japan}
\affiliation{$^3$Department of Physics, National Cheng Kung University, Tainan 701, Taiwan}
\affiliation{$^4$National Center for Theoretical Sciences, Hsinchu 300, Taiwan}
\affiliation{$^5$Department of Physics, University of Michigan, Ann Arbor, Michigan 48109-1040 USA}

\begin{abstract}
Einstein-Podolsky-Rosen (EPR) steering describes how different ensembles of quantum
states can be remotely prepared by measuring one particle of an entangled
pair. Here, we investigate quantum
steering for \emph{single} quantum $d$-dimensional systems (qudits) and devise efficient conditions to certify the steerability therein, which are applicable both to single-system steering \emph{and} EPR steering. In the single-system case our
steering conditions enable the unambiguous ruling-out of
generic classical means of mimicking steering. Ruling out `false-steering' scenarios has implications for
securing channels against both cloning-based individual attack and coherent attacks when implementing
quantum key distribution using qudits. We also show that these steering conditions also have applications in quantum computation, in that they
can serve as an efficient criterion for the evaluation of quantum logic gates of arbitrary size. Finally, we describe how the non-local EPR variant of these conditions also function as tools for identifying  faithful one-way quantum computation, secure entanglement-based quantum communication, and genuine multipartite EPR steering.
\end{abstract}

\pacs{03.65.Ud, 03.67.Dd, 03.67.Lx}
\maketitle

\section{Introduction}

Einstein-Podolsky-Rosen (EPR) steering was originally
introduced by Schr\"{o}dinger \cite{Schrodinger36} in response to the EPR
paradox \cite{EPR35}. Such steering is the ability of one party, Alice, to affect the state of another remote party, Bob, through her choice of measurement \cite{Schrodinger36}. This relies on
both the entanglement of the pair shared between Alice and Bob and the
measurement settings chosen for each particle of the pair. Recently, the concept of EPR steering has been
reformulated in terms of a information-theoretic task \cite{Wiseman07} showing that two parties
can share entanglement even if the measurement devices of one of them are
uncharacterized (or untrusted). This new formulation also illustrates a strict
hierarchy between Bell non-locality, steering and entanglement. It is worth
noting that, like Bell inequalities and entanglement witnesses, which have
been widely used to verify quantum correlations, EPR steering inequalities \cite{Cavalcanti09} and steering measures \cite{Skrzypczyk14} are introduced to detect the steerability of bipartite quantum systems. Several experimental demonstrations of EPR steering have been reported \cite{Saunders10,Smith12,Wittmann12}. Furthermore,  the steering effect has application to quantum key distribution (QKD)
when one of the parties can not trust their measurement apparatus, i.e., one-sided device-independent QKD (1SDI-QKD) \cite{Branciard12}.

Since the reformulation of EPR steering by Wiseman {\em et al.} \cite{Wiseman07}, there has been a range of investigations into steering's unique properties, quantification and potential extensions. For example, it has been shown that there exist entangled states by which steering can in only one direction \cite{Midgley10,Olsen13,Bowles14}, from Alice to Bob but not from Bob to Alice. In addition, the original bipartite steering effect has been generalized to genuine multipartite steering \cite{He13,Armstrong14,Cavalcanti14,Li15}. Moreover, a temporal analog of the steering inequality has been introduced \cite{Chen14}, and a nontrivial operational meaning to violations of such an inequality was found through a connection to the security bounds of QKD \cite{Chen14}.

\begin{figure*}[t]
\includegraphics[width=8 cm,angle=0]{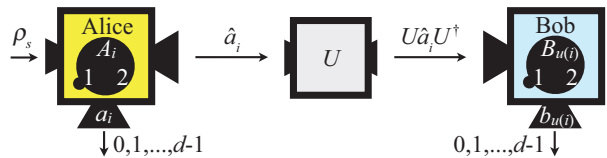}
\caption{Single-system steering for quantum information tasks. The state $\hat{a}
_{i}$ is sent from Alice to Bob. Here $\hat{a}_{i}$ is a
post-measurement state of a qudit $\protect\rho _{s}$ under the measurement $A_{i}$ for $i=1,2$. By sharing certain information distributed via a
classical communication channel (not shown), Alice can steer the state of Bob's particle by asking him to perform the quantum operation $U$. For example, by
simply choosing $U$ as an identity operator, Alice's steering enables them
to realize QKD. When $U$ is an arbitrary quantum logic
gate, steering single systems is equivalent to performing quantum
computation. To identify whether Alice can implement such steering, Bob can
use the steering condition~(\ref{wdu}) or~(\ref{wenp}) to rule out the results
mimicked by generic classical strategies. As illustrated, Bob
performs measurements $B_{u(i)}$ to implement these certifications. These
steering conditions ensure secure quantum communication and faithful quantum
computation (see table \ref{table}). Here, it is allowed that Alice and Bob have no spatial separation but
access the single system at different times.}\label{qsss}
\end{figure*}

With this greatly advanced understanding of quantum steering, a natural question arises: does there exist a strict and experimentally efficient criteria for quantum steering that can be used to certify the reliability of \emph{both} quantum communication and quantum computation tasks? So far, it was been shown that 1SDI-QKD \cite{Branciard12} benefits from EPR steering. However, there is no unified scheme for the use of quantum steering for generic quantum information processing. In fact, the role of quantum steering in quantum computation, if any, is not clear.

Here, we present a simple but unified picture to connect quantum steering with generic quantum information tasks. See Fig.~\ref{qsss} for a schematic illustration of a typical implementation. Two novel steering conditions are introduced to identify genuine single-system
quantum steering in the presence of errors and which can be applied to both quantum computation and quantum communication using qudits (systems of arbitrary dimension). Both steering conditions need only the \emph{minimum} of two local measurement settings for experimental implementation. Our results give a strict meaning of violating the temporal analog of the steering inequality \cite{Chen14} and extend the 1SDI-QKD from qubit \cite{Branciard12} to qudit cases. Moreover, we show how these conditions can be applied in the standard non-local EPR setting and then used to validate quantum computation for both the quantum circuit model \cite{Nielsen00} and one-way quantum computing \cite{OnewayQC}. Finally, we discuss the implications for certifying genuine multipartite EPR steering and implementing multipartite secret sharing with partial uncharacterized measurement devices.

\section{Quantum steering for single systems}

In the scenario of single-system quantum steering, Alice's ability to affect the quantum state
Bob has access to is based on both her ability to prepare an arbitrary quantum state to send to Bob and her knowledge, if any, about
the state Bob finally receives (which may differ from her prepared state, for various reasons) \cite{normalEPR}. If Alice has full information
about the quantum system Bob is holding, she is capable of \emph{steering} this system into an arbitrary state. Alice can
follow two steps to achieve this (Fig.~\ref{qsss}).

First, Alice prepares a specific state of a qudit with a given initial state
 state $\rho _{s}$ generated from some quantum source, before sending it to Bob,
by performing complementary measurements $A_{i}$ for $i=1,2$. Once the particle
is measured with a chosen $A_{i}$, $\rho _{s}$ becomes  $\hat{a}
_{i}\equiv \left| a_{i}\right\rangle_{ii}\!\left\langle a_{i}\right| $ for $a_{i}\in
\mathbf{v}=\{0,1,...,d-1\}$, where the $d$ states constitutes an orthonormal
basis $\{\left| a_{i}\right\rangle_{i}\}$ \cite{normalEPR}. The set of states $\{\left|
a_{2}\right\rangle_{2}\}$ is complementary to the state set $\{\left|
a_{1}\right\rangle_{1}\}$ by defining $\left| a_{2}\right\rangle_{2} =1/\sqrt{d
}\sum_{a_{1}=0}^{d-1}\omega ^{a_{2}a_{1}}\left| a_{1}\right\rangle_{1}$, with $\omega =\exp(i2\pi /d)$.

Second, the particle in the state $\hat{a}_{i}$ is then sent to Bob. Here Bob will not know the state of
particle $\hat{a}_{i}$ sent from Alice. To steer Bob's state $\hat{a}_{i}$ into other quantum states $\mathcal{U}(\hat{a}_{i})\equiv U\hat{a}_{i}U^{\dag }$, Alice can directly perform the unitary operation $U$ by herself before the particle transmission, or publicly, via a classical channel, ask Bob to apply $U$ on $\left|a_{i}\right\rangle_{i}$. While the quantum operation $\mathcal{U}$ is announced publicly, the state $\mathcal{U}(\hat{a}_{i})$ is
still unknown to Bob. It is clear that Alice has complete knowledge about the quantum system held by Bob
since the state $\rho _{s}$, the measurement $A_{i}$ and the subsequent
operation $\mathcal{U}$ are designed by Alice. When Bob performs
measurements on his particle after the operation $\mathcal{U}$, his two
complementary measurements $B_{u(i)}$ for $i=1,2$ are specified by the
orthonormal bases $\{\left| b_{u(i)}\right\rangle_{u(i)} \equiv U\left|
b_{i}\right\rangle_{i} |b_{u(i)}=b_{i}\in \mathbf{v}\}$ with the results $\{b_{u(i)}\}$.

In an ideal case, the state received by Bob is the same as the initial
state $\hat{a}_{i}$ prepared by Alice under the transformation $\mathcal{U
}$. In practical situations, however, noise from the
environment or other artificial effects introduce an unknown source of randomness. In order to explicitly
qualify whether Alice can steer the states of the particles eventually held
by Bob, and rule out either third-party eavesdropping, classical mimicry of the channel, or to qualify the quality of the channel itself,
we consider the following generic \emph{classical means}
of describing state preparation, transitions between states, and the limits to which they can influence the measurement results of Bob.

First, we assume that the state of the particle sent by Alice can be described by a classical realistic theory which
predicts the particle is in a state described by a fixed set $(A_{1}=a_{1},A_{2}=a_{2})$. Suppose next that $P(a_{1},a_{2})$ is the probability that, before the measurements are performed, the particle is in a state $(a_{1},a_{2})$. Under this assumption the marginal probability $P(a_{i})$ and the conditional probability $P(a_{i}|a_{j})$ for $i,j=1,2$ and $i\neq j$ should follow the relation
\begin{equation}
P(a_{1},a_{2})=P(a_{1})P(a_{2}|a_{1})=P(a_{2})P(a_{1}|a_{2}).  \label{gre}
\end{equation}
Second, we assume that the particle state can change, while it is being transmitted from Alice to Bob, from $(a_{1},a_{2})$ to an unknown state $\rho _{\lambda }$ with a transition probability $P[\lambda |(a_{1},a_{2})]$. Then, the state of the particle changes to $\sum_{a_{1},a_{2}}P(a_{1},a_{2})\sum_{\lambda}P[\lambda |(a_{1},a_{2})]\rho _{\lambda }$. To connect this state with our steering scenario, where the state of the particle, and how it evolves, may depend on the choice to measure $a_{1}$ or $a_{2}$ individually, we rewrite the transition probability as $P[\lambda |(a_{1},a_{2})]=P(\lambda|a_{i})P(a_{j}|\lambda,a_{i})/P(a_{j}|a_{i})$ \cite{TranProp}. From which, combined with the relation~(\ref{gre}), the joint probability of finding $(a_{1},a_{2})$ and observing $\lambda$ as the final state can be explicitly represented by
\begin{eqnarray}
P[(a_{1},a_{2}),\lambda ]&=&P(a_{1},a_{2})P[\lambda |(a_{1},a_{2})]\nonumber\\
&=&P(a_{i})P(\lambda |a_{i})P(a_{j}|\lambda ,a_{i}).  \label{ret}
\end{eqnarray}
As shown by (\ref{gre}) and (\ref{ret}), it does not matter what order Alice does a series of measurements, the joint probability will always be the same. The state of the particle that Bob holds is then
\begin{equation}
\rho _{B}=\sum_{a_{i}=0}^{d-1}P(a_{i})\sum_{\lambda }P(\lambda|a_{i})\rho _{\lambda }.  \label{rb}
\end{equation}
When summing over all $a_{1}$ and $a_{2}$, Eq.~(\ref{ret}) becomes
\begin{equation}
P(\lambda )=\sum_{a_{1}}P(a_{1})P(\lambda |a_{1})=\sum_{a_{2}}P(a_{2})P(\lambda |a_{2}).  \label{re0}
\end{equation}
With the above classical realistic description of Alice's states, the state received
by Bob becomes independent of the measurement setting chosen by Alice, i.e., $\rho _{B}=\sum_{\lambda }P(\lambda )\rho _{\lambda }$, implying that Bob always has the same state whatever measurement $A_{i}$ and operation $\mathcal{U}$ Alice designs. This means Alice cannot steer Bob's states. We call the states with this feature \emph{unsteerable}. The above proof can be seen as equivalent to that used in the derivation of EPR steering inequalities and extended EPR steering conditions, where Alice's measurement results are assumed to be a classical distribution. See Appendix A for detailed discussions.

Finally, if Alice's state and the unknown states $\rho_{\lambda}$ are
described by a classical theory of realism, and thus only classically correlated with Bob's results, then the descriptions Eqs. (\ref{gre}), (\ref{ret}) and (\ref
{re0}) are applicable to $\rho_{\lambda}$ as well. However, here Bob's measurement results are assumed to be based on measurements on a quantum particle. Thus the expectation values of
the two mutually-unbiased measurements $B_{u(1)}$ and $B_{u(2)}$ with
respect to the unknown quantum states $\rho_{\lambda}$ obey the quantum uncertainty
relation in the entropic form \cite{Tomamichel11}
\begin{equation}
H(B_{u(1)}|\lambda)+H(B_{u(2)}|\lambda)\geq \log_{2}(d),  \label{uncertain}
\end{equation}
where $H(B_{u(i)}|\lambda)=-\sum_{b_{u(i)}=0}^{d-1}P(b_{u(i)}|\lambda)
\log_{2}P(b_{u(i)}|\lambda)$.

\section{Quantum steering conditions}

\subsection{Steering conditions}
In order to distinguish steerability from the results mimicked by the methods based on the classical
theories considered above, in what follows we will introduce two novel
quantum steering conditions of the from $\mathcal{S}>\alpha_{R}$, where $\mathcal{S}$ is the kernel of the criterion and $\alpha_{R}$ is the maximum value of the kernel supported by classical theories. For ideal steering, $\mathcal{S}$ will be maximized. Since ruling out classical mimicry is equivalent to excluding unsteerable states (\ref{rb}), exceeding the $\alpha_{R}$ will deny, or rule out, processes (e.g., noisy channels) that make once steerable states unsteerable and thus assist in confirming genuine quantum steering.

The kernel of our first steering condition is
\begin{equation}
\mathcal{S}_{dU}\equiv \sum_{i=1}^{2}\sum_{a_{i}=0;b_{u(i)}=a_{i}}^{d-1}P(a_{i},b_{u(i)}).  \label{w}
\end{equation}
For ideal steering the maximum value for the kernel is $\mathcal{S}_{dU}=2$. Whereas, for the
states described by Eq.~(\ref{rb}), we have $\alpha _{R}=1+1/\sqrt{d}$. Thus the quantum steering condition reads
\begin{equation}
\mathcal{S}_{dU}>1+\frac{1}{\sqrt{d}}.  \label{wdu}
\end{equation}
For any unsteerable states the measured kernel will not violate this bound. To determine the maximum value of the kernel supported by realistic theories, we consider the expectation value of the kernel $\mathcal{S}_{dU}$ for the state $\rho _{B}$ (\ref{rb}). Then $\mathcal{S}_{dU}$ becomes
\begin{equation}
\mathcal{S}_{dU,R}=\sum_{i=1}^{2}\sum_{a_{i}=0}^{d-1}\sum_{\lambda }\text{Tr}[U\left|a_{i}\right\rangle_{ii\!}\left\langle a_{i}\right|U^{\dagger}\rho _{\lambda }]P(\lambda |a_{i})P(a_{i}).\nonumber
\end{equation}
This can be further manipulated to give
\begin{eqnarray}
\mathcal{S}_{dU,R}&\leq& \sum_{\lambda }P(\lambda )\big(\text{Tr}[\left|m\right\rangle_{11\!}\left\langle m\right|\rho _{\lambda }]+\text{Tr}[\left|n\right\rangle_{22\!}\left\langle n\right|\rho _{\lambda }]\big)\nonumber\\
&\leq&1+\frac{1}{\sqrt{d}},\nonumber
\end{eqnarray}
where $m,n\in \mathbf{v}$. The first inequality is derived by using the relation (\ref{re0}) about $P(\lambda)$, and the classical bound $\alpha _{R}=1+1/\sqrt{d}$ is then obtained by determining the maximum eigenvalue of the operator $\left|m\right\rangle_{11\!}\left\langle m\right|+\left|n\right\rangle_{22\!}\left\langle n\right|$.

Our second steering condition is based on the mutual information between Alice
and Bob. From the point of view of information shared between sender and
receiver, the ability for Alice to steer Bob's state is confirmed if the mutual dependence
between the measurement results of Alice and Bob is stronger than the
dependence of Bob's measurement outcomes on the unknown states $
\rho_{\lambda}$ and $\rho_{B}$. This condition of
steerability can be represented in terms of the mutual information as follows,
\begin{equation}
\sum_{i=1}^{2}I(B_{u(i)};A_{i})>\sum_{i=1}^{2}I(B_{u(i)};\{\lambda\}).  \label{entro0}
\end{equation}
From the basic definition of mutual information, Eq.~(\ref{entro0}) implies
that
\begin{equation}
\sum_{i=1}^{2}\sum_{a_{i}=0}^{d-1}P(a_{i})H(B_{u(i)}|a_{i})<\sum_{i=1}^{2}\sum_{
\lambda}P(\lambda)H(B_{u(i)}|\lambda).\nonumber
\end{equation}
Imposing the relation (\ref{uncertain}) on the state $
\rho_{\lambda}$, we obtain the second steering condition of the form
\begin{equation}
\mathcal{S}_{\rm {ent}\textit{U}}=-\sum_{i=1}^{2}\sum_{a_{i}=0}^{d-1}P(a_{i})\, H(B_{u(i)}|a_{i})>\log_{2}\left(\frac{1}{d}\right).  \label{wenp}
\end{equation}

In addition to the steering conditions devised here, violating the temporal
steering inequality \cite{Chen14} can serve as an indicator of single-system
steering. In Appendix B, we show that this inequality can
be derived from the classical conditions~(\ref{gre}) and~(\ref{rb}), which provides a strict meaning of violating that inequality. As shown therein, the steering conditions are related to practical quantum information tasks and then more useful than the temporal steering inequality from a practical point of view. See Appendix C for a concrete demonstration of the sensitivity of these conditions.

In particular, one of the main advantages of the steering criteria is that they can be efficiently implemented in experiments. The \emph{minimum} two
measurement settings are sufficient to measure the kernels $\mathcal{S}_{dU}$ and $\mathcal{S}_{\rm {ent}\textit{U}}$. In addition, they are robust against noise. See Appendix D for demonstrations of the robustness of our steering conditions and the EPR steering inequality for single systems.

\subsection{Implications of the steering conditions}

We use the generic classical means of describing state preparation and transitions between states to consider the threshold $\alpha_{R}$ for the steering conditions. Such conditions certify quantum steering (EPR steering and single-system steering) when the measurement apparatus of Alice is uncharacterized or both of the Alice's measurement devices and the operation $U$ are untrusted.

It is important to note that ruling out these classical mimicries is equivalent to excluding the unsteerable states (\ref{rb}). Thus satisfying these conditions will deny, or rule out, processes that make states unsteerable. For example, it is possible that, while the measurement devices of Alice functions as well as expected, any processes that can change the states of particles from $\hat{a}_{1}$ and $\hat{a}_{2}$ to unknown states belonging to $\{\rho_{\lambda}\}$ will cause Alice to be ignorant about the true connection between her true measurement outcomes and Bob's states. Such state changes make Bob's state unsteerable as described by Eq. (\ref{rb}).

In practical situations, one usually does not know the full information about the noise from the
environment, or other artificial effects which introduce an unknown source of randomness. The steering conditions (\ref{wdu}) and (\ref{wenp}) can certify the ability of Alice to steer the states of the particles eventually held
by Bob, and then rule out third-party eavesdropping, classical mimicry of the channel and any processes that make the transmitted particles unsteerable. Hence these steering conditions can be considered as an objective tool to evaluate the reliability of quantum communication and quantum computation.

\section{Quantum communication}

When the state of the qudit sent from Alice to Bob changes from the
state $\hat{a}_{i}$ to a state $\mathcal{U}_{\rm real}(\hat{a}_{i})$ through a
channel $\mathcal{U}_{\rm real}$, the value of the kernel $\mathcal{S}_{dU}$ is 
\begin{equation}
\mathcal{S}_{dU}=\sum_{i=1}^{2}\sum_{a_{i}=0}^{d-1}P(a_{i})\, F(a_{i},u(i)),\nonumber
\end{equation}
 where the probabilities $P(a_{i})=\text{Tr}[\rho _{s}\hat{a}_{i}]$ and the state fidelities \cite{Nielsen00} $
F(a_{i},u(i))=\text{Tr}[\mathcal{U}_{\rm real}(\hat{a}_{i})\hat{a}_{u(i)}]$. Let us assume that an error is introduced
by a quantum cloning machine \cite{Cerf02} which copies equally well the states of both bases, $F(a_{i},u(i))=F$, for all $a\in \mathbf{v}$ \cite{Fidelity}. If Alice wants to demonstrate steering of Bob's particle in the presence of such
eavesdropping, they have to find $\mathcal{S}_{dU}=2F>1+1/\sqrt{d}
$, or alternatively the state fidelity must satisfy the condition:
\begin{equation}
F>\frac{1}{2}(1+\frac{1}{\sqrt{d}}).\nonumber
\end{equation}
 It is equivalent to saying that the disturbance, $D=1-F$, or error rate,
has to be lower than a certain upper bound $D_{\text{ind}}=(1-1/\sqrt{d})/2$. This bound is exactly the same as the well known security threshold \cite{Cerf02}.

For the second steering condition~(\ref{wenp}), we derive a second criterion on the state fidelity $F$ \cite{Fidelity}: 
\begin{equation}
\tilde{F}>-\frac{1}{2}\log _{2}(d),\nonumber
\end{equation}
where $\tilde{F}\equiv F\log _{2}(F)+(1-F)\log _{2}\left[(1-F)/(d-1)\right]$. This provides the upper bound, $D_{\text{coh}}$, on $D$ under coherent attacks. If
$D<D_{\text{coh}}$, then Alice can steer Bob's state. Interestingly, such derived $D_{
\text{coh}}$ exactly coincides with the existing result \cite
{Cerf02,Sheridan10}. The above two conditions on $F$ are summarized in table \ref{table}.

\section{Quantum computation}

When the measured kernels $\mathcal{S}$ are larger
than the maximum values $\alpha_{R}$ predicted by classical theories, the
real process describing the state transitions $\mathcal{U}_{\rm real}$ can be said to be
close to the target unitary quantum operations $\mathcal{U}$ that Alice and Bob expect \cite{Computation}. To consider how to evaluate such a transformation further, we rewrite
the condition (\ref{wdu}) as 
\begin{equation}
\frac{1}{d}\sum_{i=1}^{2}\sum_{a_{i}=0}^{d-1}\text{Tr}[\mathcal{U}_{\rm real}(\hat{a}
_{i})\hat{a}_{u(i)}]>1+\frac{1}{d}.\nonumber
\end{equation}
Here, without losing any generality, we assume that $\rho_{s}=I/d$, where $I$
is the identity matrix. The quantity
\begin{equation}
F_{\hat{a}_{i}\rightarrow \mathcal{U}(\hat{a}_{i})}\equiv \frac{1}{d}\sum_{a=0}^{d-1}\text{Tr}[\mathcal{U}_{\rm real}(\hat{a}_{i})\hat{a}_{u(i)}]\nonumber
\end{equation}
can be considered as an average fidelity between $\mathcal{U}_{\rm real}(\hat{a}_{i})$ and $\hat{a}_{u(i)}$ over all the $d
$ states. With the average state fidelities $F_{\hat{a}_{i}\rightarrow
\mathcal{U}(\hat{a}_{i})}$ for the complementary bases $A_{1}$ and $A_{2}$,
one can obtain the lower bound of the process fidelity $F_{\text{process}}\equiv
\text{Tr}[\mathcal{U}_{\rm real}\mathcal{U}]$ by $F_{\text{
process}}\geq F_{\hat{a}_{1}\rightarrow \mathcal{U}(\hat{a}_{1})}+F_{\hat{a}
_{2}\rightarrow \mathcal{U}(\hat{a}_{2})}-1$ \cite{Hofmann05}. Hence, using the steering condition together with the above relation, we obtain a condition for a faithful quantum process in terms of process fidelity: 
\begin{equation}
F_{\text{process}}>\frac{1}{d}. \nonumber
\end{equation}

Taking a two-qubit entangling gate for an example, this indicator coincides with the well known criterion \cite{Hofmann05} in terms of the concurrence $C$ \cite{Hill97}.  Two qubits can be considered or recast as a single system with a level number $d=2^{2}=4$. The entanglement capability of a two-qubit entangling gate, like a controlled-{\sc not} operation, can be defined by the minimal amount of entanglement that can be generated by the real operation $\mathcal{U}_{rel}$. In terms of the concurrence $C$, a measure of quantum entanglement, it is found that $C\geq 2F_{\text{process}}-1$ \cite
{Hofmann05}. Then, for a nontrivial gate, one requires $C>0$, which implies that $F_{\text{process}}>1/2$. Our condition on $F_{\text{process}}$ derived from the steering condition (\ref{wdu}) coincides with this criterion. Note that the condition derived from the second steering condition (\ref{wenp}) is $F_{\text{process}}>62.14\%$ and tighter than that resulted from the condition (\ref{wdu}).

The above results can be efficiently implemented with the minimum two
measurement settings. This is especially useful to evaluate experimental quantum logic gates of arbitrary size, for example, an experimental three-qubit Toffoli gate with trapped ions \cite{Monz09}. For a three-qubit gate ($d=2^{3}$), the condition on the process fidelity is $F_{\text{process}}>1/\sqrt{8}\approx 35.36\%$. The process fidelity of the experimental quantum Toffoli gate with trapped ions reported in \cite{Monz09} is $F_{\text{process}}=66.6(4)\%$, which can be identified as being functional according to our proposed criterion. When the number of qubits $N$ increases, the classical bound will decrease with $\sqrt{d}=2^{N/2}$ and approach zero when $N$ is
large. 

The second steering condition~(\ref{wenp}) can be used to evaluate
experimental quantum gates. When using the same
conditions as $D_{\text{coh}}$ to consider the quality of gate operations
under coherent attacks, one can obtain the condition on $F_{\text{process}}$ in
terms of $D_{\text{coh}}$:
\begin{equation}
F_{\text{process}}>1-2D_{\text{coh}},\nonumber
\end{equation}
which is tighter than the criterion derived from the first condition (\ref{wdu}). The
relation $F=F_{\hat{a}_{1}\rightarrow \mathcal{U}(\hat{a}_{1})}=F_{\hat{a}
_{2}\rightarrow \mathcal{U}(\hat{a}_{2})}$ is used above. Alternatively, the
gate can be also qualified if the average state fidelity
satisfies $F>1-D_{\text{coh}}$. Table \ref{table} summarizes the above two conditions on $F_{\text{process}}$.

\begin{table}
\caption{\label{table}A summary of the steering conditions for quantum information processing. The criteria derived from steering conditions for secure quantum communications and faithful quantum computations are represented in terms of the state fidelity $F$ and the process fidelity $F_{\text{process}}$, respectively.}
\begin{ruledtabular}
\begin{tabular}{lll}
Condition  & Communication & Computation\\
  $\mathcal{S}_{dU}>1+\frac{1}{\sqrt{d}}$ & $F>\frac{1}{2}\left(1+\frac{1}{\sqrt{d}}\right)$ & $F_{\text{process}}>\frac{1}{\sqrt{d}}$\\
  $\mathcal{S}_{\rm ent\textit{U}}>\log_{2}\left(\frac{1}{d}\right)$ & $\tilde{F}>-\frac{1}{2}\log _{2}(d)$ &$F_{\text{process}}>1-2D_{\text{coh}}$
\end{tabular}
\end{ruledtabular}
\end{table}

\section{EPR steering conditions and applications}

As discussed above, traditional EPR-steering and single-system-steering scenarios mirror each other. In the language we use, this can be understood from the fact that, by changing the role of $\lambda$ \cite{lambdaEPR}, both steering conditions~(\ref{wdu}, \ref{wenp}) can be used to detect EPR steering for \emph{bipartite} $d$-level systems shared between Alice and Bob. See Appendix A.1.d. However, the converse is also true, such that EPR steering inequalities, for example, the inequalities used in the experiments \cite{Saunders10,Smith12}, can serve as criteria for single-system steering (see Ref. \cite{Chen14} and Appendix B).

When using the bipartite counterpart of steering conditions~(\ref{wdu}, \ref{wenp}) for quantum communication, one obtains security criteria for quantum channels that are the same as the single-system case, which can thus be considered as a $d$-level extension of 1SDI-QKD \cite{Branciard12}. Similarly, the EPR steering conditions give criteria of computation performance for quantum gates realized in one-way modes \cite{OnewayQC}. A quantum gate $U$ can be encoded in a bipartite maximally-entangled state \cite{cluster}:
\begin{equation}
\left|U\right\rangle=\frac{1}{\sqrt{d}}\sum_{a_{i}=0}^{d-1}\left|a_{i}\right\rangle_{i}\left|\text{Out}(a_{i})\right\rangle,\nonumber
\end{equation}
 where $\left|\text{Out}(a_{i})\right\rangle\equiv U\left|\text{In}(a_{i})\right\rangle$, and $\left|\text{In}(a_{i})\right\rangle$ is the input state of the quantum gate $U$. A readout of the gate operation, $\left|\text{Out}(a_{i})\right\rangle$, depends on the measurement result $a_{i}$, which is just the effect of EPR steering. See Appendix E for an application to a two-qubit gate realized in the one-way mode. Hence our EPR steering conditions can indicate reliable gate operations for experiments \cite{OnewayQCexp} in the presence of uncharacterized measurement devices.

The idea of bipartite steering conditions based on~(\ref{wdu}, \ref{wenp}) can be straightforwardly generalized to genuine multipartite EPR steering. The main ingredient is to consider a kernel, from either the joint probabilities like Eq.~(\ref{w}) or the entropic conditions in  Eq.~(\ref{wenp}),  for a specific bipartition of a multipartite system. Then a complete kernel of steering condition is composed of the joint probabilities, or entropic conditions, for all possible bipartitions of the multipartite system. See \cite{Li15} for concrete examples for steering conditions based on~(\ref{wdu}). In particular, the entropic condition for genuine multipartite EPR steering using (\ref{wenp}) could be useful for multipartite quantum secret sharing \cite{Hillery99} when coherent attacks occur in the quantum network.

\section{Conclusion and outlook}

We investigated the concept of quantum
steering for single quantum systems and pointed out its role in quantum
information processing. We derived two novel steering conditions to certify such
steering. These conditions ensure secure QKD using
qudits and provide new criteria for efficiently evaluating experimentally
quantum logic gates of arbitrary computing size (see table \ref{table}). Moreover, the bipartite counterparts of our steering conditions can detect EPR steerability of bipartite $d$-level systems, and have practical uses for evaluating one-way quantum computing and quantum communication with entangled qudits and verifying genuine multipartite EPR steering. It may be interesting to
investigate further the connection between single-system steering and
other types of quantum steering such as one-way steering \cite
{Midgley10,Olsen13,Bowles14}.

\acknowledgements

 C.-M.L. acknowledges the partial support from the Ministry of Science and Technology, Taiwan, under Grant No. MOST 101-2112-M-006-016-MY3 and MOST 104-2112-M-006 -016-MY3. Y.-N.C. is partially supported by the Ministry of Science and Technology, Taiwan, under Grant No. MOST 103-2112-M-006-017-MY4.

\begin{figure*}[t]
\includegraphics[width=17.5 cm,angle=0]{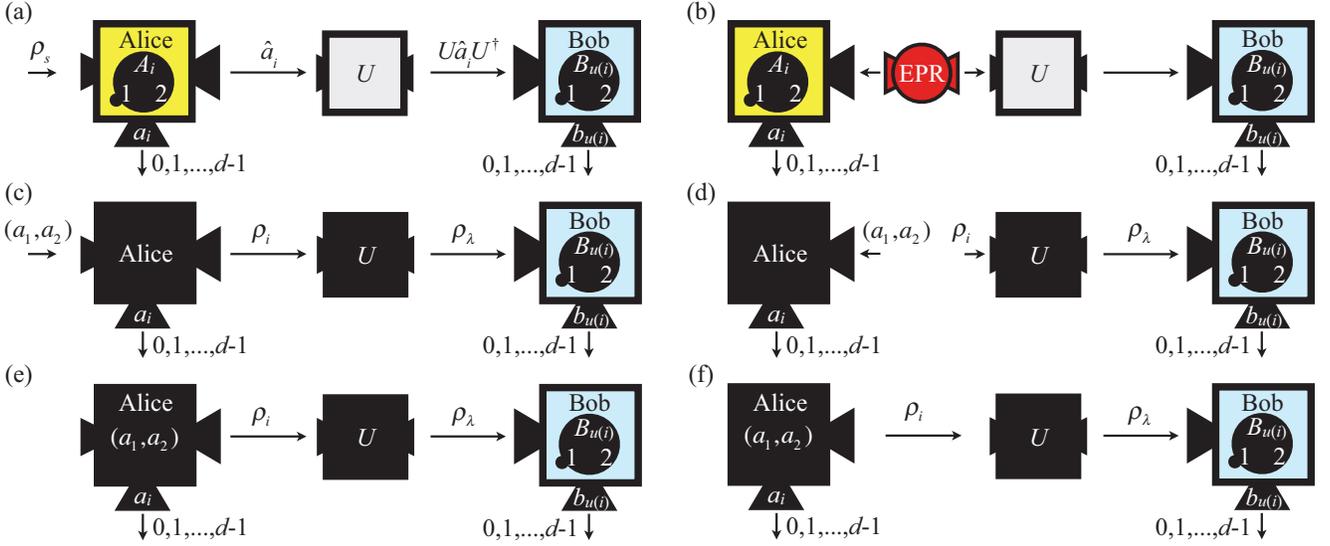}
\caption{Comparison between single-system steering and EPR steering. We compare these two scenarios by, first, their basic concepts of ideal single-system steering (a) and ideal EPR steering (b), and, second, classical mimics of single-system steering (c), (e) and EPR steering (d), (f). For the ideal case, Alice can use the effect of EPR steering, by sharing the entangled states (EPR source), to implement the operation $U$ on the state of Bob's qudit. While the resources utilized for quantum steering are different, the state of the particle finally held by Bob can be steered into a corresponding quantum state, $U\hat{a}_{i}U^{\dag }$, for both quantum steering scenarios. To distinguish classical mimicry from genuine quantum steering, the respective classical models based on realistic theories (c) and (d) are introduced. These ``classical simulations'' can be concretely represented in the practical descriptions of, for example, unqualified measurements of Alice and the unqualified operation $U$ performed by Alice or Bob [(e) and (f)]. As shown in (e) and (f), these effective simulations are equivalent.}\label{EPRSingle}
\end{figure*}

\appendix

\section{Comparing single-system steering with EPR steering}

In this section we compare EPR steering with single-system steering by discussing their basic assumptions and the classical mimicries, or simulation, of steering effects (Fig. \ref{EPRSingle}). This provides a clear connection between EPR and single-system steering and the steering conditions for both cases discussed in our work. From this comparison, we show that classical mimicry or simulation can in both cases be considered as equivalent.

\subsection{EPR steering for quantum information processing (QIP)}

Compared with the single-system steering [Fig. \ref{EPRSingle}(a)], the scenario of EPR steering also consists of two steps: First, Alice generates a bipartite entangled system from an entanglement source (or called EPR source) [Fig. \ref{EPRSingle}(b)]. To have a concrete comparison, let us assume that this entangled state is of the form
\begin{equation}
\left|\Phi\right\rangle=\frac{1}{\sqrt{d}}\sum_{a_{1}=b_{1}=0}^{d-1}\left|a_{1}\right\rangle_{A1}\otimes\left|b_{1}\right\rangle_{B1}
\end{equation}
where $\{\left|a_{1}\right\rangle_{A1}\equiv\left|a_{1}\right\rangle_{1}|a_{1}\in\textbf{v}\}$ and $\{\left|b_{1}\right\rangle_{B1}\equiv\left|b_{1}\right\rangle_{1}|b_{1}\in\textbf{v}\}$.

Second, Alice keeps one particle of the entangled pair and sends the other particle to Bob. A subsequent unitary operator $U$ is applied on the Bob's subsystem according to the instructions of Alice. This transformation can be done either by Bob after receiving the particle, or by Alice herself before the transmission of the particle. After such transformation, the state vector of the bipartite system becomes
\begin{equation}
(I\otimes U)\left|\Phi\right\rangle=\frac{1}{\sqrt{d}}\sum_{a_{1}=b_{1}=0}^{d-1}\left|a_{1}\right\rangle_{A1}\otimes U\left|b_{1}\right\rangle_{B1}.\nonumber
\end{equation}
Then, depending on Alice's measurement result $a_{1}$, the state of the particle finally held by Bob can be steered into a corresponding quantum state, $U\hat{a}_{1}U^{\dag }$, which is the same as the result derived from single-system steering. When the state $\left|\Phi\right\rangle$ is represented in the bases $\{\left|a_{2}\right\rangle_{A2}\equiv\left|a_{2}\right\rangle_{2}|a_{2}\in\textbf{v}\}$ and $\{\left|b_{2}\right\rangle_{B2}\equiv\left|b_{2}\right\rangle_{2}|b_{2}\in\textbf{v}\}$, we have
\begin{equation}
\left|\Phi\right\rangle=\frac{1}{\sqrt{d}}\sum_{a_{2}+b_{2}\doteq 0}\left|a_{2}\right\rangle_{A2}\otimes\left|b_{2}\right\rangle_{B2},
\end{equation}
where $\doteq$ denotes equality modulo $d$. Through the same method as that shown above, Alice can steer the state of Bob's particle into the quantum state, $U\hat{b}_{2}U^{\dag }$, by the measurement on her subsystem with a result $a_{2}$ satisfying the correlation $a_{2}+b_{2}\doteq 0$.

We remark that, for an EPR source creating entangled states that are different from $\left|\Phi\right\rangle$, the transformation $U$ could be implemented in other ways. For example, when Alice and Bob share bipartite supersinglets \cite{supersinglets}, which are expressed as
\begin{equation}
\left|\Psi\right\rangle=\frac{1}{\sqrt{d}}\sum_{a_{i}+b_{i}=d-1}(-1)^{a_{i}}\left|a_{i}\right\rangle_{Ai}\otimes\left|b_{i}\right\rangle_{Bi},
\end{equation}
for $i=1,2$, Alice can steer the state of Bob by directly measuring her qudit in a basis featured in $U$. Since supersinglets are rotationally invariant \cite{supersinglets}, i.e., $(R\otimes R)\left|\Psi\right\rangle=\left|\Psi\right\rangle$, where $R$ is a rotation operator, Alice's measurement in the basis $\{R\left|a_{i}\right\rangle_{i}\}$ will steer the state of Bob's qudit into a corresponding state, $R\left|b_{i}\right\rangle_{i}$, for $a_{i}+b_{i}=d-1$. For $d=2$, supersinglets become unitary invarient and provide a resource for implementing any unitary transformations $U$ to Bob's qubit.

\subsection{Steering conditions}

For both ideal single-system and EPR steering scenarios, the state received by Bob, $\hat{b}_{i}$, is the same as or perfectly correlated with the initial state $\hat{a}_{i}$ prepared by Alice under the transformation $U$. However, for Bob's limited knowledge about the measurements used or the particle prepared by Alice, her measurement results become untrusted to Bob. He is uncertain whether these measurements and state preparation are qualified. In the worst case where Alice's measurement outcomes may be randomly generated from her apparatus, classical simulations then can describe Alice's measurement results. To show that Alice has true steerability in practical situations, the steering conditions (\ref{wdu}) and (\ref{wenp}) have been introduced to distinguish genuinely quantum steering from the classical mimicry. In what follows, we will detail the classical mimicry and their implication for practical applications. With these examples, it will be clear that the proof for single-system conditions can be seen as equivalent to that used in the derivation of EPR steering conditions.

\subsubsection{Mimicry of single-system steering}

In the case of single-system steering, as detailed in the main text, the classical mimicry of steering is based on the realistic assumptions that (1) the state of the particle sent by Alice can be described by a fixed set $(a_{1},a_{2})$, and (2) the state can change from $(a_{1},a_{2})$ to another state $\lambda$ which corresponds to a quantum state of the qudit $\rho_{\lambda}$ finally held by Bob, see Fig.~\ref{EPRSingle}(c). In order to see this mimicry from a practical point of view, one can think that, for example, such a situation arises as a result of the unqualified measurement device and the states of particles sent to Bob. For some reason, Alice's measurement apparatus does not properly output real measurement results $a_{i}$ but randomly generates outcomes with a distribution $P(a_{1},a_{2})$ [see Eq.~(\ref{gre})] that correspond to some output states $\rho_{i}$ as the measurement setting $i$ is chosen by Alice. After the unqualified operation  $U$, the state $\rho_{i}$ becomes the unknown state $\rho_{\lambda}$ which constitutes an unsteerable state $\rho_{B}$ [Eq. (\ref{rb})]. Here the joint probability of finding $(a_{1},a_{2})$ and observing $\lambda$ as the final state satisfies the classical relation (2). It is equivalent to say that Alice can consider the joint set $(a_{1},a_{2})$, with the probability of occurrence $P(a_{1},a_{2})$, as describing predetermined instructions for her to prepare and send a particle with final quantum state $\rho_{\lambda}$ to Bob. See Fig.~\ref{EPRSingle}(e).

\begin{figure*}[t]
\includegraphics[width=17.5 cm,angle=0]{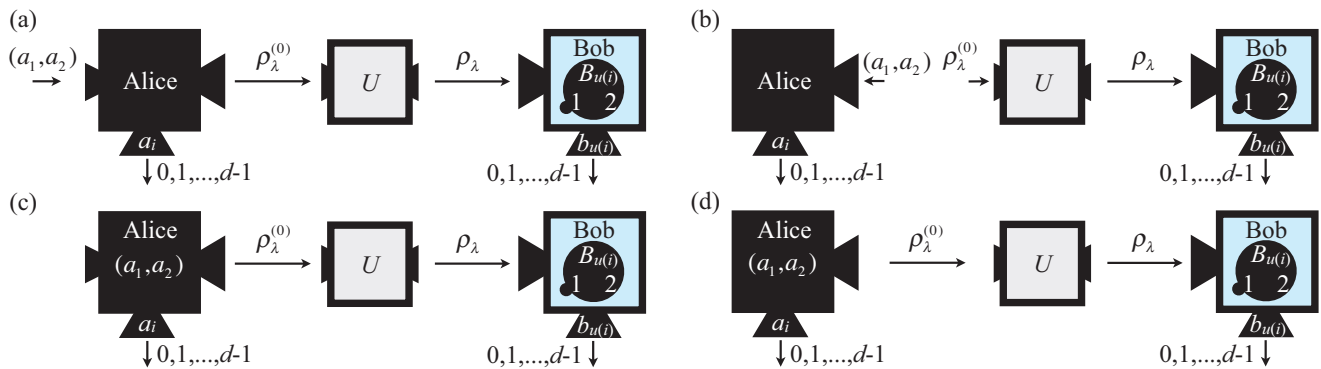}
\caption{Steering mimicries where the operation $U$ is qualified but the measurement device of Alice is not. The classical mimics of single-system steering (a) and EPR steering (b) are based on the realistic assumptions that (1) the state of the particle sent by Alice can be described by a fixed set $(a_{1},a_{2})$, and (2) the state can change from $(a_{1},a_{2})$ to another state $\lambda$ which corresponds to a quantum state of the qudit $\rho_{\lambda}$ finally held by Bob. One can concretely represent these scenarios in the practical descriptions of unqualified Alice's apparatus (c) and (d), respectively. These concrete simulations are then shown to be equivalent.}\label{EPRSingle2}
\end{figure*}

It is also possible that the operation $U$ is qualified but the measurement device of Alice is not. The two realism assumptions are applicable to this case as well. The above classical mimicry scenario can be recast such that the output states $\rho_{i}$ already correspond to the unknown state $\rho^{(0)}_{\lambda}$, see Fig.~\ref{EPRSingle2}(a). It does not matter what the subsequent qualified operation on the particle $U$ is, the final states held by Bob $\rho_{\lambda}$ constitute an unsteerable state $\rho_{B}$. From a practical point of view, similarly, one can think that Alice's measurement apparatus randomly generates outcomes with the probability of occurrence $P(a_{1},a_{2})$ that correspond to unknown output states $\rho^{(0)}_{\lambda}$ [Fig.~\ref{EPRSingle2}(c)].

\subsubsection{Mimicry of EPR steering}

The above scheme for mimicking single-system steering can be readily mapped to the case of EPR steering. Here, the mimicry of EPR steering depends on two similar assumptions: (1) the state of the particle held by Alice can be described by a fixed set obeying realism $(a_{1},a_{2})$, and (2) a given set $(a_{1},a_{2})$ corresponds to some quantum state, $\rho_{\lambda}$, of the qudit finally held by Bob, see Fig.~\ref{EPRSingle}(d). The unqualified bipartite state shared between her and Bob, and a subsequent unqualified operation, can result in such assumptions. For example, let us assume that the entanglement source does not create entangled pairs but a qudit with state $\rho_{i}$ for Bob and another separable particle for Alice instead. For the state $\rho_{i}$ there is a corresponding measurement setting $i$ chosen by Alice, for which Alice's measurement device creates an output of a random signal with a distribution described by the probability $P(a_{1},a_{2})$ (\ref{gre}). The subsequent operation $U$ takes $\rho_{i}$ to an unknown state $\rho_{\lambda}$, and then the final state held by Bob is unsteerable (\ref{rb}). The classical relation (\ref{ret}) is again applicable to this transition between states. Here it is reasonable to incorporate the entanglement source into the measurement apparatus as a single unqualified experiment setup for Alice. See Fig.~\ref{EPRSingle}(f). Then it is effectively a scenario where Alice observes a set $(a_{1},a_{2})$ appearing with probability $P(a_{1},a_{2})$ which creates a particle with a final quantum state $\rho_{\lambda}$ for Bob.

As discussed in the above mimicry of single-system steering, it is possible that the operation $U$ is qualified but Alice's measurement apparatus, including the EPR source, is not. In this case one can effectively consider that the unqualified EPR source outputs a fixed set $(a_{1},a_{2})$ for Alice's particle and a qudit that is already in an unknown state $\rho_{i}=\rho^{(0)}_{\lambda}$ for Bob [Fig.~\ref{EPRSingle2}(b)]. For any qualified operation  $U$ on the particle state $\rho^{(0)}_{\lambda}$, the final state held by Bob is still unsteerable. From the same practical point of view as introduced above, we can think that the joint set $(a_{1},a_{2})$, with the probability of occurrence $P(a_{1},a_{2})$, resulting from the random outcomes of Alice's device, corresponds to a particle with final quantum state $\rho_{\lambda}$ for Bob, see Fig.~\ref{EPRSingle2}(d). It is clear that the joint probability of finding $(a_{1},a_{2})$ and observing $\lambda$ as the final state in this case satisfies the classical relation (\ref{ret}).

\subsubsection{The equivalence between the steering mimicries}

With the above concrete explanations of the classical mimicry for both the single-system steering and EPR steering, one can interpret these two classical scenarios as being equivalent to each other. See (e) and (f) in Fig.~\ref{EPRSingle} and (c) and (d) in Fig.~\ref{EPRSingle2}. Following the same approach based on the realistic assumptions and their practical scenarios, in what follows we will discuss two more cases to complete the proof of the equivalence between the steering mimicries.

\begin{figure*}[t]
\includegraphics[width=17.5 cm,angle=0]{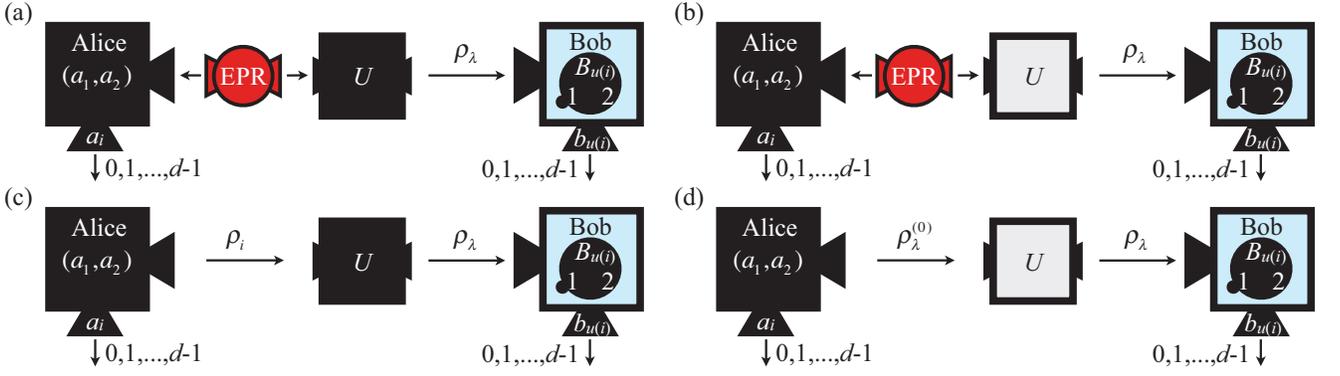}
\caption{Mimicries of EPR steering where the EPR source is qualified but the measurement device of Alice is not. (a) the unqualified operation is used, and (b) the operation used is qualified in the mimicry. These two possible situations can be described by the two realism conditions and represented in practical descriptions (c) and (d), respectively. The demonstration (c) has the analogue of single-system steering described by Fig.~\ref{EPRSingle}(e). The concrete mimicry (d) is equivalent to that of single-system steering depicted in Fig.~\ref{EPRSingle2}(c).}\label{EPRSingle3}
\end{figure*}

The case where Alice's measurement apparatus is unqualified, while the EPR source functions as expected, can raise two other  possible scenarios which again can be shown to be covered by "realism" assumptions. Figure~\ref{EPRSingle3}(a) depicts one of the possibilities. As the operation $U$ is unqualified, one can practically think that Alice's measurement apparatus generates random outcomes with a distribution $P(a_{1},a_{2})$, independent of the entangled pair generated from the EPR source. The subsequent operation makes the state of the qudit of the entangled pair sent to Bob, say $\rho_{i}$, change to $\rho_{\lambda}$ as illustrated by Fig.~\ref{EPRSingle3}(c). It is clear that such mimicry of EPR steering is equivalent to the simulation of single-system steering described by Fig.~\ref{EPRSingle}(e) [see also Fig.~\ref{EPRSingle}(c)].

Figure~\ref{EPRSingle3}(b) illustrates the other situation where the entanglement source and the operation $U$ are qualified but Alice's measurement apparatus is not. One of concrete examples for this case is as the following. The unqualified measurement device of Alice always measures her particle of the entangled pair, say $\left|\Phi\right\rangle$, in the first basis $\{\left|a_{1}\right\rangle_{1}|a_{1}\in\textbf{v}\}$ intrinsically whatever measurement setting Alice chooses, and it announces random signals $a_{1}$ or $a_{2}$ as an outcome. Such Alice's measurement and the random signals announced make the state of the qudit sent to Bob unsteerable, i.e, $\rho_{i}=\rho_{\lambda}^{(0)}$ belongs to the same set $\{\left|b_{1}\right\rangle_{1}|b_{1}\in\textbf{v}\}$ whatever measurement setting chosen by Alice and as such then constitutes an unsteerable state $\rho_{B}$ after the operation $U$. See Fig.~\ref{EPRSingle3}(d). This is an analogue of EPR-steering mimicry to that of single-system steering described by Fig.~\ref{EPRSingle2}(c) [see also Fig.~\ref{EPRSingle2}(a)].

\subsubsection{EPR steering conditions for QIP}

As shown above, the mimicry of single-system steering is equivalent to that mimicking EPR steering. Then the steering conditions (\ref{wdu}) and (\ref{wenp}) for single systems can be mapped to verifications of EPR steering for bipartite $d$-dimensional systems. All such EPR steering conditions can certify the reliability of QIP when entangled pairs are shared between Alice and Bob. For the EPR steering conditions which correspond to the criterion (7), when the state $\left|\Phi\right\rangle$ is used to mediate steering the condition is of the form
\begin{eqnarray}
\mathcal{S}^{(\text{EPR})}_{dU\Phi}&\equiv& \sum_{a_{1}=b_{u(1)}=0}^{d-1}P(a_{1},b_{u(1)})\nonumber\\
&&+\sum_{a_{2}+b_{u(2)}\doteq 0}P(a_{2},b_{u(2)})>1+\frac{1}{\sqrt{d}}.\label{seprphi}
\end{eqnarray}
Similarly, with proper changes to the above joint probabilities, we have the following steering condition for supersinglets
\begin{eqnarray}
\mathcal{S}^{(\text{EPR})}_{dR\Psi}&\equiv& \sum_{a_{u(1)}+b_{u(1)}=d-1}P(a_{u(1)},b_{u(1)})\nonumber\\
&&+\!\!\!\sum_{a_{u(2)}+b_{u(2)}=d-1}\!\!\!\!\! \!\!\!\!\! P(a_{u(2)},b_{u(2)})>1+\frac{1}{\sqrt{d}},
\end{eqnarray}
where, for Alice who implements quantum measurements, her measurement outcomes $\{a_{u(i)}\}$  result from the measurement described by the basis $\{\left| a_{u(i)}\right\rangle_{u(i)} \equiv R\left|
a_{i}\right\rangle_{i} |a_{u(i)}=a_{i}\in \mathbf{v}\}$. Here Bob uses the same measurements as that used by Alice. For the EPR steering conditions represented in the entropic forms, we have
\begin{equation}
\mathcal{S}^{(\text{EPR})}_{\rm ent\textit{U}\Phi}\equiv-\sum_{i=1}^{2}\sum_{a_{i}=0}^{d-1}P(a_{i})\, H(B_{u(i)}|a_{i})>\log_{2}\left(\frac{1}{d}\right),
\end{equation}
for the state $\left|\Phi\right\rangle$ shared by Alice and Bob, and
\begin{equation}
\mathcal{S}^{(\text{EPR})}_{\rm ent \textit{R}\Psi}\equiv-\sum_{i=1}^{2}\sum_{a_{u(i)}=0}^{d-1}P(a_{u(i)})\, H(B_{u(i)}|a_{i})>\log_{2}\left(\frac{1}{d}\right),
\end{equation}
for the supersinglets.

As detailed above, the mimicry of single-system steering based on realistic theories is equivalent to that of EPR steering where Alice's outcomes follows realist theories but Bob performs quantum measurements. Hence the proof for the conditions (\ref{wdu}) and (\ref{wenp}) can be readily applied to the above EPR steering conditions. In addition, following the same analysis of quantum communication based on single-system steering as introduced in the main text, these bipartite counterpart of steering conditions provide security criteria for quantum channels that is equivalent to the single-system cases.

\section{EPR steering inequality for single-system steering}

The classical condition~(\ref{gre}) and the results derived from which such as Eqs. (\ref{ret}) and (\ref{rb}) provide a strict meaning of violating the single-system analogue of the EPR steering inequality used in the experiment of Smith \textit{et al.} \cite{Smith12}, i.e., the temporal steering inequality introduced in \cite{Chen14}. The kernel of such steering inequality reads
\begin{equation}
S_{N}\equiv \sum_{i=1}^{N}E[\left\langle
B_{i,t_{B}}\right\rangle^{2}_{A_{i,t_{A}}}],
\end{equation}
where
\begin{equation}
E[\left\langle B_{i,t_{B}}\right\rangle^{2}_{A_{i,t_{A}}}]\;=\;
\sum_{a=0}^{1}P(A_{i,t_{A}}=a)\;\left\langle
B_{i,t_{B}}\right\rangle^{2}_{A_{i,t_{A}}=a}
\end{equation}
and $N=2$ or $3$ is the number of measurement for Alice and Bob. The
probability of measuring $A_{i}=a$ at the time $t_{A}$ is denoted by $
P(A_{i,t_{A}}=a)$. The expectation value about Bob's measurement at the time
$t_{B}$, conditioned on the measurement result of Alice, is defined by
\begin{eqnarray}
\left\langle
B_{i,t_{B}}\right\rangle_{A_{i,t_{A}}=a}&=&
\sum_{b=0}^{1}(-1)^{b}\;P(B_{i,t_{B}}=b|A_{i,t_{A}}=a).  \notag
\end{eqnarray}
To obtain the upper bound derived from generic classical means, we firstly
introduce the final state of Bob's particle~(\ref{rb}) into the above equation and
then have
\begin{eqnarray}
\left\langle
B_{i,t_{B}}\right\rangle_{A_{i,t_{A}}=a}\;&=&\;\sum_{b=0}^{1}(-1)^{b}\!\sum_{
\lambda}\!P(B_{i,t_{B}}=b|\lambda)P(\lambda|A_{i,t_{A}}=a)  \notag \\
&=&\;\sum_{\lambda}P(\lambda|A_{i,t_{A}}=a)\;\left\langle
B_{i,t_{B}}\right\rangle_{\lambda}.\notag
\end{eqnarray}
Then it is clear that
\begin{eqnarray}
&&E[\left\langle B_{i,t_{B}}\right\rangle^{2}_{A_{i,t_{A}}}]\nonumber\\
&\leq&\sum_{a=0}^{1}P(A_{i,t_{A}}=a)\sum_{\lambda}P(\lambda|A_{i,t_{A}}=a)\;\left
\langle B_{i,t_{B}}\right\rangle^{2}_{\lambda}.\nonumber
\end{eqnarray}
Secondly, we use the result (\ref{re0}) derived from the criterion on state transition~(\ref{ret}) in the main text to obtain
\begin{equation}
P(\lambda)=\sum_{a=0}^{1}P(A_{i,t_{A}}=a)P(\lambda|A_{i,t_{A}}=a),\nonumber
\end{equation}
for all measurements $i$. The temporal inequality is
\begin{equation}
S_{N}\leq\sum_{i=1}^{N}\sum_{\lambda}P(\lambda)\left\langle
B_{i,t_{B}}\right\rangle^{2}_{\lambda}\leq\sum_{\lambda}P(\lambda)=1.\nonumber
\end{equation}
Thus $S_{N}>1$ can be considered as a condition for single-system steering and deny processes that make states unsteerable.

\section{Comparison between steering conditions and the temporal steering inequality}

One of the main difference between the steering conditions and the temporal steering inequality is in their practical applications to quantum information tasks. In what follows we will illustrate a simple example to show that, compared with the temporal steering inequality, the steering conditions can fulfil certain requirements so as to useful as checks for the reliability of QIP.

Let us assume that a source generates particles in the state $\rho_{s}=\left|0\right\rangle_{11}\left\langle 0\right|$ for Alice's subsequent use for steering. The task of Alice and Bob is to perform an identity operation $I$, or alternatively, to maintain the states of the particles during the particle transmission. For such an information task, the steering condition (7) for $d=2$ and $U=I$ used by them to check the steerability can be of the form
\begin{equation}
\mathcal{S}_{2I}\equiv\sum_{a_{1}=b_{1}=0}^{1}P(a_{1},b_{1})+\sum_{a_{2}=b_{2}=0}^{1}P(a_{2},b_{2})>1+\frac{1}{\sqrt{2}}.\nonumber
\end{equation}
When the particles are transmitted without any disturbance, they will have $\mathcal{S}_{2I}=2$. To concretely show the undesired situation, e.g., a wrong gate operation in quantum computation, or an unwanted interaction between the qubit and the quantum channel inquantum communication, we assume that there exists an effective operation $X=\left|0\right\rangle_{11\!}\left\langle 1\right|+\left|1\right\rangle_{11\!}\left\langle 0\right|$ on the qubit such that the final state of the qubit held by Bob is $X\left|b_{i}\right\rangle_{i}$. Such an operation can make the qubit flip when the state is prepared in $\left|0\right\rangle_{1}$ or $\left|1\right\rangle_{1}$. Then the value of the kernel $\mathcal{S}_{2I}$ becomes $\mathcal{S}_{2I}=1$, i.e., the reliability of maintaining qubit is not certified by the steering condition (7).

Whereas, using the same number of measurement settings ($N=2$), the temporal steering inequality is still violated by $S_{N}=2$, and this can not reveal the real effect of a qubit flip on the particle during transmission. Hence, the present form of the temporal steering inequality can not be used in practical quantum information tasks. However, after properly revising the kernel $S_{N}$ by introducing a quantum operation $U$ for quantum communication or quantum computation, the revised version of the temporal inequality also can serve the same role as the steering conditions. Its derivation and experimental demonstrations will be detailed elsewhere.

The above consideration is also true for the bipartite non-local counterpart. When Alice and Bob share the state $\left|\Phi\right\rangle=\frac{1}{\sqrt{2}}\sum_{a_{1}=b_{1}=0}^{1}\left|a_{1}\right\rangle_{A1}\otimes\left|b_{1}\right\rangle_{B1}$ to perform the same task as above, they can certify the reliability by using the steering condition (\ref{seprphi}) for $d=2$ and $U=I$
\begin{equation}
\mathcal{S}^{(\text{EPR})}_{dU\Phi}\equiv \sum_{a_{1}=b_{1}=0}^{1}P(a_{1},b_{1})+\sum_{a_{2}=b_{2}=0}^{1}P(a_{2},b_{2})>1+\frac{1}{\sqrt{2}}.\nonumber
\end{equation}
If there is a bit flip error in the transmission of Bob's qubit, then the state suffering from such effect $(I\otimes X)\left|\Phi\right\rangle$ can not give results that satisfy the above condition to act as a reliability check ($\mathcal{S}^{(\text{EPR})}_{dU\Phi}=1$) but still can violate the inequality ($S_{N}=2>1$). Then, the EPR steering inequality can not respond to the effect of a qubit flip in the bipartite non-local scenario.

\begin{figure}[t]
\includegraphics[width=8.5 cm,angle=0]{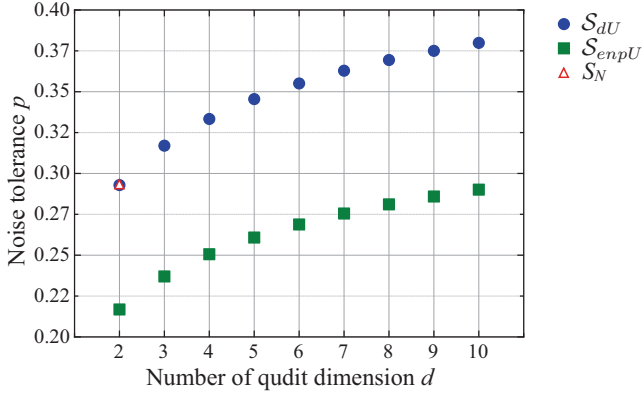}
\caption{Noise tolerance of steering conditions (\ref{wdu}) and (\ref{wenp}). If the probability of white nose $p_{\text{noise}}<p$, then the single-system steering, that underlies the qudits sent by Alice with the states $\rho_{i}(a_{i},p_{\text{noise}})$, can be certified by the steering conditions (\ref{wdu}) or (\ref{wenp}). Here the threshold of noise intensity $p$ is an indicator showing the noise tolerance of these steering criteria. Note that the noise tolerance of the certification based on violating the EPR steering inequality, i.e., $S_{N}>1$, implemented with two measurement settings ($N=2$) is the same as that of the steering condition (\ref{wdu}) for two-dimensional systems (the EPR steering inequality introduced by Smith \textit{et al.} \cite{Smith12} is applicable to $d=2$ only). For large $d$, both the conditions (\ref{wdu}) and (\ref{wenp}) are robust against noise up to $p=50\%$. }\label{robust}
\end{figure}

\section{Robustness of steering conditions}

We consider the following scenario to determine the robustness of the proposed steering conditions. Let us suppose that in the presence of white noise the pure state $\left|a_{i}\right\rangle_{i}$ of the qudit prepared by Alice's measurements will become
\begin{equation}
\rho_{i}(a_{i},p_{\text{noise}})=\frac{p_{\text{noise}}}{d}I+(1-p_{\text{noise}})\hat{a}_{i},
\end{equation}
where $p_{\text{noise}}$ is the probability of uncolored noise. Then the steerability revealed by using the qudits with states $\rho_{i}(p_{\text{noise}})$ is certified by our steering conditions if the intensity of uncolored noise $p_{\text{noise}}$ is small than some noise threshold, $p_{\text{noise}}<p$. Here $p$ can be considered as an indicator showing the noise tolerance of the steering conditions. See Fig.~\ref{robust}. We determine the noise threshold $p$ by considering the critical noise intensity such that $\mathcal{S}(p_{\text{noise}})=\alpha_{R}$. For the steering condition (\ref{wdu}), we have
\begin{equation}
p=\frac{(1-\frac{1}{\sqrt{d}})}{2(1-\frac{1}{d})},
\end{equation}
which shows that the steering condition is robust and the noise is even tolerable up to $p=50\%$ for large $d$. The robustness of the steering condition (\ref{wenp}) is similar to that of the condition (\ref{wdu}), and its noise tolerance in terms $p$ also can be up to $p=50\%$ for large $d$.

\begin{figure*}[t]
\includegraphics[width=12 cm,angle=0]{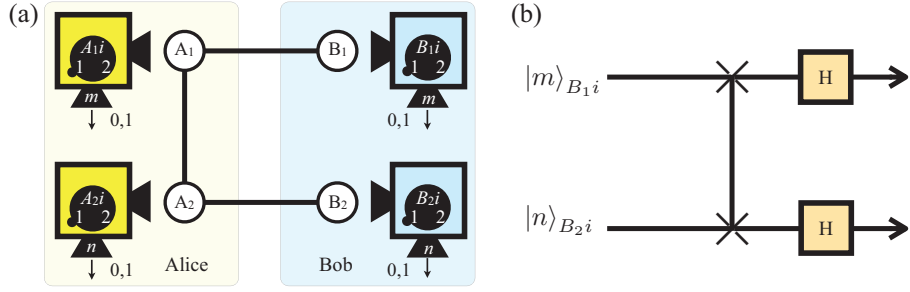}
\caption{EPR steering for one-way quantum computing. (a) A genuine four-qubit chain-type cluster state shared by Alice and Bob is represented by a fully-connected horseshoe graph. Alice, who performs the measurements $A_{1}i$ and $A_{2}i$ for $i=1,2$ on her qubits $\text{A}_{1}$ and $\text{A}_{2}$, respectively, can reveal the EPR steering effect to realize the gate operation $U$ on the qubits of Bob $\text{B}_{1}$ and $\text{B}_{2}$. (b) The state $\left|m\right\rangle_{B_{1}i}\otimes\left|n\right\rangle_{B_{2}i}$ is an input of the quantum gate $U$ composed of one two-qubit {\sc cphase} gate and two single-qubit Hadamard operations. For one-way quantum computing, the outcomes of Alice's measurements, $m$ and $n$, corresponding to the post measurement state $\left|m\right\rangle_{A_{1}i}\otimes\left|n\right\rangle_{A_{2}i}$, determines the output state of the gate operation, $U\big(\left|m\right\rangle_{B_{1}i}\otimes\left|n\right\rangle_{B_{2}1i}\big)$.  }\label{OnewayQC}
\end{figure*}

\section{EPR steering for one-way quantum computing}

A cluster state can be represented by an array of vertices, where
each vertex is initially in the state of $
\left( \left| 0\right\rangle +\left| 1\right\rangle
\right)/\sqrt{2}$ where $\left|0\right\rangle$ and $\left|1\right\rangle$ constitutes an orthonormal basis. Every connected line (edge) between vertices realises
a controlled-phase ({\sc cphase}) gates acting as $\left|
m\right\rangle\otimes\left| n\right\rangle \rightarrow \omega
^{mn}\left| m\right\rangle\otimes\left| n\right\rangle$, where $\omega=\exp(i2\pi/2)$ and $m,n\in
\left\{ 0,1\right\}$ \cite{OnewayQC}. In the present illustration, we consider a four-qubit chain-type cluster state of the form
\begin{equation}
\left|C_{4}\right\rangle=\!\!\sum_{m=0}^{1}\sum_{n=0}^{1}\sum_{j=0}^{1}\sum_{k=0}^{1}\omega^{mn+nj+jk}\left|n\right\rangle_{A_{1}}\!\otimes\left|j\right\rangle_{A_{2}}\!\otimes\left|m\right\rangle_{B_{1}}\!\otimes\left|k\right\rangle_{B_{2}}
\end{equation}
where $\left|q\right\rangle_{A_{l}}=\left|q\right\rangle_{B_{r}}\equiv\left|q\right\rangle$ for $q=0,1$ and $l,r=1,2$. The state $\left|C_{4}\right\rangle$ represented in a horseshoe graph is shown in Fig.~\ref{OnewayQC}(a). Here we assume that Alice holds two of the qubits, $\text{A}_{1}$ and $\text{A}_{2}$, and Bob has the rest, $\text{B}_{1}$ and $\text{B}_{2}$.

When sharing such a genuine four-partite entangled state between them, Alice's quantum measurements on her qubits can realize a quantum gate operation $U$ on the state of the qubits held by Bob:
\begin{equation}
U=(H\otimes H)\text{{\sc cphase}},
\end{equation}
where $H$ is the Hadamard operation, see Fig.~\ref{OnewayQC}(b). To clearly see the gate operation realized in this one-way model, we rephrase the state vector of $\left|C_{4}\right\rangle$ in the following form
\begin{equation}
\left|C_{4}\right\rangle=\sum_{m=0}^{1}\sum_{n=0}^{1}\left|m\right\rangle_{A_{1}1}\otimes\left|n\right\rangle_{A_{2}1}\otimes U\big(\left|m\right\rangle_{B_{1}1}\otimes\left|n\right\rangle_{B_{2}1}\big)
\end{equation}
where $\left|q\right\rangle_{A_{l}1}=\left|q\right\rangle_{B_{r}1}\equiv(\left|0\right\rangle+(-1)^{q}\left|1\right\rangle)/\sqrt{2}$ for $q=0,1$ and $l,r=1,2$. One can consider the state $\left|m\right\rangle_{B_{1}1}\otimes\left|n\right\rangle_{B_{2}1}$ as an input of the quantum gate $U$. Then the outcomes of Alice's measurements $A_{1}1$ and $A_{2}1$, $m$ and $n$, corresponding to the post measurement state $\left|m\right\rangle_{A_{1}1}\otimes\left|n\right\rangle_{A_{2}1}$, determines the output state of the gate operation, $U\big(\left|m\right\rangle_{B_{1}1}\otimes\left|n\right\rangle_{B_{2}1}\big)$. For example, as Alice performs measurements and has the results $m=0$ and $n=0$, the state of Bob's qubits $(\left|0\right\rangle+\left|1\right\rangle)\otimes(\left|0\right\rangle+\left|1\right\rangle)/2$ will be transformed by $U$ into an entangled state $(\left|0\right\rangle\otimes\left|0\right\rangle+\left|0\right\rangle\otimes\left|1\right\rangle+\left|1\right\rangle\otimes\left|0\right\rangle-\left|1\right\rangle\otimes\left|1\right\rangle)/2$. Alice can perform different measurements to transform input states prepared in different basis by the same gate operation $U$. The cluster state also can be of the form
\begin{equation}
\left|C_{4}\right\rangle=\sum_{m=0}^{1}\sum_{n=0}^{1}\left|m\right\rangle_{A_{1}2}\otimes\left|n\right\rangle_{A_{2}2}\otimes U\big(\left|m\right\rangle_{B_{1}2}\otimes\left|n\right\rangle_{B_{2}2}\big)
\end{equation}
where $\left|q\right\rangle_{A_{l}2}=\left|q\right\rangle_{B_{r}2}\equiv(\left|0\right\rangle+(-1)^{q}i\left|1\right\rangle)/\sqrt{2}$ for $q=0,1$ and $l,r=1,2$.

Through the connection between Alice's measurements on her qubits and the resulting states of Bob's qubits as illustrated above, one can think of the quantum gate $U$ as being encoded in a bipartite maximally-entangled state
\begin{equation}
\left|U\right\rangle=\frac{1}{2}\sum_{a_{i}=0}^{3}\left|a_{i}\right\rangle_{i}\otimes\left|\text{Out}(a_{i})\right\rangle,
\end{equation}
where $\left|a_{i}\right\rangle_{i}\equiv \left|m\right\rangle_{A_{1}i}\otimes\left|n\right\rangle_{A_{2}i}$ with $a_{i}=m\times 2^{1}+n\times 2^{0}$ and $\left|\text{Out}(a_{i})\right\rangle\equiv U\left|\text{In}(a_{i})\right\rangle$, and $\left|\text{In}(a_{i})\right\rangle\equiv \left|m\right\rangle_{B_{1}i}\otimes\left|n\right\rangle_{B_{2}i}$ is the input state of the quantum gate $U$. Hence the effect of EPR steering reveals that a readout of the gate operation, $\left|\text{Out}(a_{i})\right\rangle$, depends on the measurement result $a_{i}$.

Our EPR steering conditions serves an useful tool to identify reliable gate operations for experiments in the presence of uncharacterized (or untrusted) measurement devices. For example, for the above concrete case, we have the following EPR steering conditions
\begin{eqnarray}
\mathcal{S}^{(\text{EPR})}_{dUC_{4}}&\equiv& \sum_{a_{1}=b_{u(1)}=0}^{3}P(a_{1},b_{u(1)})\nonumber\\
&&+\sum_{a_{2}=b_{u(2)}=0}^{3}P(a_{2},b_{u(2)})>3/2. \label{onewayss}
\end{eqnarray}
where $\{b_{u(i)}\}$ denotes the results obtained from Bob's measurement specified by $\{\left| b_{u(i)}\right\rangle_{u(i)} \equiv U\left|\text{In}(b_{i})\right\rangle |b_{u(i)}=b_{i}\in \mathbf{v}\}$. It is easy to find that the kernel $\mathcal{S}^{(\text{EPR})}_{dUC_{4}}$ and its condition for EPR steering are exactly the same as their single-system analogues (\ref{w}) and (\ref{wdu}).

It is worth noting that the idea of bipartite EPR steering effects and the steering condition (\ref{onewayss}) for one-way quantum computing is rather different from that based on genuine multipartite EPR steering \cite{Li15}. The present steering condition detects EPR steering with respect to the fixed bipartite splitting of the four qubits $\text{A}_{1},\text{A}_{2}$ and $\text{B}_{1},\text{B}_{2}$. When certifying \emph{genuine} four-partite EPR steering for one-way quantum computing, one needs the concept and method introduced in \cite{Li15} to consider and verify quantum steering with respect to all bipartite splittings of the four qubits.


\begin{references}

\bibitem{Schrodinger36} E. Schr\"odinger, Proc. Cambridge Philos. Soc. \textbf{31}, 553 (1935); \textbf{32}, 446 (1936).

\bibitem{EPR35} A. Einstein, B. Podolsky, and N. Rosen, Phys. Rev. \textbf{47}, 777 (1935).

\bibitem{Wiseman07} H. M. Wiseman, S. J. Jones, and A. C. Doherty, Phys. Rev. Lett. \textbf{98}, 140402 (2007).

\bibitem{Cavalcanti09} E. G. Cavalcanti, S. J. Jones, H. M. Wiseman, and M. D. Reid, Phys. Rev. A \textbf{80}, 032112 (2009).

\bibitem{Skrzypczyk14} P. Skrzypczyk, M. Navascues, and D. Cavalcanti, Phys. Rev. Lett. \textbf{112}, 180404 (2014).

\bibitem{Saunders10}  D. J. Saunders, S. J. Jones, H. M. Wiseman, and G. J. Pryde, Nat. Phys. \textbf{6}, 845 (2010).

\bibitem{Smith12} D.-H. Smith, G. Gillett,	M. P. de Almeida, C. Branciard, A. Fedrizzi, T. J. Weinhold, A. Lita, B. Calkins, T. Gerrits, H. M. Wiseman, S. W. Nam, and A. G. White, Nat. Commun. \textbf{3}, 625 (2012).

\bibitem{Wittmann12} B. Wittmann, S. Ramelow, F. Steinlechner, N. K. Langford, N. Brunner, H. M. Wiseman, R. Ursin, and A. Zeilinger,  New J. Phys. \textbf{14}, 053030 (2012).

\bibitem{Branciard12} C. Branciard, E. G. Cavalcanti, S. P. Walborn, V. Scarani, and H. M. Wiseman, Phys. Rev. A \textbf{85}, 010301(R) (2012).

\bibitem{Midgley10} S. L. W. Midgley, A. J. Ferris, and M. K. Olsen, Phys. Rev. A \textbf{81}, 022101 (2010).

\bibitem{Olsen13} M. K. Olsen, Phys. Rev. A \textbf{88}, 051802 (2013).

\bibitem{Bowles14} J. Bowles, T. V\'ertesi, M. T. Quintino, and N. Brunner, Phys. Rev. Lett. \textbf{112}, 200402 (2014).

\bibitem{He13} Q. Y. He and M. D. Reid, Phys. Rev. Lett. \textbf{111}, 250403 (2013).

\bibitem{Armstrong14} S. Armstrong, M. Wang, R. Y. Teh, Q. Gong, Q. He, J. Janousek, H.-A. Bachor, M. D. Reid, and P. K. Lam, arXiv:1412.7212.

\bibitem{Cavalcanti14} D. Cavalcanti, P. Skrzypczyk, G. H. Aguilar, R. V. Nery, P. H. Souto Ribeiro, S. P. Walborn, arXiv:1412.7730.

\bibitem{Li15} C.-M. Li, K. Chen, Y.-N. Chen, Q. Zhang, Y.-A. Chen, J.-W. Pan, Phys. Rev. Lett. \textbf{115}, 010402 (2015).

\bibitem{Chen14} Y.-N. Chen, C.-M. Li, N. Lambert, S.-L. Chen, Y. Ota, G.-Y. Chen, and F. Nori, Phys. Rev. A \textbf{89}, 032112 (2014).

\bibitem{Nielsen00} M. A. Nielsen and I. L. Chuang, \textit{Quantum Computation and Quantum Information} (Cambridge University Press, Cambridge, England, 2000).

\bibitem{OnewayQC}  H. J. Briegel and R. Raussendorf, Phys. Rev. Lett. \textbf{86}, 910 (2001); R. Raussendorf and H. J. Briegel, \textit{ibid.} \textbf{86}, 5188 (2001).

\bibitem{normalEPR} In normal EPR steering, Alice can steer
Bob's state into arbitrary target states only when the pair of
particles are entangled and she knows the state structure of the entangled
pair shared between them. The state information enables Alice to choose a
proper measurement basis to demonstrate steering. This is the same for single-system steering.
Such an equivalence means that, with steering conditions alone, Bob cannot tell whether his
quantum system is one part of the entangled pair or a single particle pre-prepared and sent
from Alice (though a scheme can be devised to distinguish these two \cite{Chen14}, as can a case-by-case analysis of the allowed correlations between measurement results \cite{Ried15}). As with the role of entanglement played
in EPR steering, the essence of single system \emph{steerability} is the quantum characteristics of the states $\hat{a}_{i}$, for example,
quantum coherence and uncertainty relations.

\bibitem{Ried15} K. Ried, M. Agnew, L. Vermeyden, D. Janzing, R. W. Spekkens  and K. J. Resch, Nat. Phys. \textbf{11}, 414 (2015).

\bibitem{TranProp} Here we have utilized the relation $P(\lambda,a_{j}|a_{i})=P(a_{j}|a_{i})P[\lambda |(a_{1},a_{2})]=P(\lambda|a_{i})P(a_{j}|\lambda,a_{i})$. The transition probability $P[\lambda |(a_{1},a_{2})]$ is then connected with the individual transition probability $P(\lambda |a_{i})$.

\bibitem{Tomamichel11} M. Tomamichel and R. Renner, Phys. Rev. Lett. \textbf{106}, 110506 (2011).

\bibitem{Cerf02} N. J. Cerf, M. Bourennane, A. Karlsson, and N. Gisin, Phys. Rev. Lett. \textbf{88}, 127902 (2002).

\bibitem{Fidelity} If all the states before Bob's measurements $\mathcal{U}_{\rm real}(\hat{a}_{i})$ are identical to the states $\hat{a}_{u(i)}$, i.e., $F(a_{i},u(i))=1$, it is clear
that $\mathcal{S}_{dU}=2$. Whereas, if there exists an error source which reduces the state fidelity $F(a_{i},u(i))$, the value of the kernel $\mathcal{S}_{dU}$ will decrease as well. If a cloner makes all the state fidelities
under the same measurement setting have the same value, say $F(a_{1},u(1))=F$
and $F(a_{2},u(2))=\bar{F}$, for all $a\in \mathbf{v}$. Then $\mathcal{S}_{dU}$ becomes $\mathcal{S}_{dU}=F+\bar{F}$. When the cloning machine copies equally well the
states of both bases, then the state fidelities in both bases are identical, $F=\bar{F}$. For the second criterion on the state fidelity, it is worth noting that the
conditional entropy can be represented by $H(B_{u(i)}|a_{i})=-F(a_{i},u(i))\log
_{2}F(a_{i},u(i))-\sum_{b\neq a}\Omega (b_{u(i)}a_{i})\log _{2}\Omega
(b_{u(i)}a_{i})$, where $\Omega (b_{u(i)}a_{i})$ denotes the probability of
error state transition from $a_{i}$ to $b_{u(i)}$ for $b_{u(i)}\neq a_{i}$. When taking the same condition as on the quantum
cloning machine for the first criterion into consideration
and assuming that the possible errors are
equiprobable $\Omega (b_{u(i)}a_{i})=(1-F)/(d-1)$, we derive a second
criterion on the state fidelity $F$ from the second steering condition (\ref{wenp}).

\bibitem{Sheridan10} L. Sheridan and V. Scarani, Phys. Rev. A \textbf{82}, 030301(R) (2010).

\bibitem{Computation} This evaluation is based on whether the process $\mathcal{U}_{\rm real}$ goes beyond the classical descriptions of the input states and their state evolution, and gives us a tool by which to evaluate a given real transformation.

\bibitem{Hofmann05} H. F. Hofmann, Phys. Rev. Lett. \textbf{94}, 160504 (2005).

\bibitem{Hill97} S. Hill and W. K. Wootters, Phys. Rev. Lett. \textbf{78}, 5022 (1997).

\bibitem{Monz09} T. Monz, K. Kim, W. Hansel, M. Riebe, A. S. Villar, P. Schindler, M. Chwalla, M. Hennrich, and R. Blatt, Phys. Rev. Lett. \textbf{102}, 040501 (2009).

\bibitem{lambdaEPR} One can change the role of $\lambda$ from that of variables for describing correlations between Bob and Alice's results via unknown states to hidden random variables for describing correlations between Alice's classical state and Bob's quantum one.

\bibitem{cluster} A one-way quantum computer relies on genuine multipartite cluster states \cite{OnewayQC} to perform gate operations. Here the state $\left|U\right\rangle$ for one-way quantum computing is the Schmidt form of cluster states with respect to a fixed bipartition, which splits the total systems into measurement part and readout of quantum gate. The Schmidt rank $d$, of the state $\left|U\right\rangle$, then represents the size of computation. For example \cite{OnewayQCexp} (see also Appendix E), a four-qubit cluster state can be used to implement quantum circuit composed of two-qubit gates, and its Schmidt rank is $d=4$ for such a bipartition.

\bibitem{OnewayQCexp} P. Walther, K. J. Resch, T. Rudolph, E. Schenck, H. Weinfurter, V. Vedral, M. Aspelmeyer and A. Zeilinger, Nature (London) \textbf{434}, 169 (2005); K. Chen, C.-M. Li, Q. Zhang, Y.-A. Chen, A. Goebel, S. Chen, A. Mair, and J.-W. Pan, Phys. Rev. Lett. \textbf{99}, 120503 (2007).

\bibitem{Hillery99} M. Hillery, V. Bu\v{z}ek, and A. Berthiaume, Phys. Rev. A \textbf{59}, 1829 (1999).

\bibitem{supersinglets} A. Cabello, Phys. Rev. Lett. \textbf{89}, 100402 (2002); A. Cabello, J. Mod. Opt. \textbf{50}, 10049 (2003).


\end{references}
\end{document}